\documentclass[prb, preprint, amsmath, amssymb, showpacs, superscriptaddress,longbibliography]{revtex4-1}
\usepackage{graphicx} 
\usepackage{dcolumn}  
\usepackage{bm}       
\usepackage{amsfonts}
\usepackage{dsfont}
\usepackage{csquotes}

\usepackage{ulem}
\usepackage{color}

\usepackage{float}
\usepackage{physics}

\usepackage{hyperref}

\usepackage[dvipsnames]{xcolor} 
\usepackage{easyReview}
\begin{document}

\title{Photocurrent generation in solids via linearly polarized laser}

\author{Amar Bharti}
\affiliation{%
Department of Physics, Indian Institute of Technology Bombay,
           Powai, Mumbai 400076, India }

\author{Gopal Dixit}
\email[]{gdixit@phy.iitb.ac.in}
\affiliation{%
	Department of Physics, Indian Institute of Technology Bombay,
	Powai, Mumbai 400076, India }
\affiliation{%
Max-Born Institut, Max-Born Stra{\ss}e 2A, 12489 Berlin, Germany }

\date{\today}

\pacs{}

\begin{abstract}
To add to the rapidly progressing field of ultrafast photocurrent, we propose a universal method to generate photocurrent in normal and topological materials using a pair of multicycle linearly polarized laser pulses.
The interplay of the fundamental and its second harmonic pulses is studied
for the generation of photocurrent in Weyl semimetals 
by varying the angle between the polarization direction, relative intensity, and relative phase delay.
It has been found that the  presence of a comparatively weaker second harmonic pulse is sufficient to generate  
substantial photocurrent. 
Moreover, significant photocurrent is generated even when polarization directions are 
orthogonal for  certain ratios of the lasers' intensities. 
In addition, the photocurrent is found to be susceptible to the delay between the two pulses. 
We have illustrated that all our findings are extendable to non-topological and two-dimensional materials, such as   graphene and  molybdenum disulfide.
\end{abstract}

\maketitle

\section{Introduction}
Ongoing efforts to synthesize novel materials for power-efficient and fast-responding photodetectors, solar cells, and optoelectronic devices have captured our attention in recent years~\cite{Pusch_2023, ma2023photocurrent,liu2020semimetals,huo2018recent}. 
The discovery of graphene, together with transition-metal dichalcogenides, has heralded photodetection sensitivity on 
an  ultrafast time domain~\cite{koppens2014photodetectors,wang2015ultrafast,tielrooij2015generation,agarwal_2023,chen2022control,prechtel2012time,gan2013chip,xia2009ultrafast,yoshioka2022ultrafast,fang2020mid}. 
Furthermore, the latest additions to novel quantum materials, such as Dirac and Weyl semimetals, 
have fostered the prospect for efficient conversion of light to electricity~\cite{liu2020semimetals,Tian_2022,Wang_2022,Lai_2022,braun2016ultrafast,wang2017ultrafast}. 
The topologically protected states in Weyl semimetals can facilitate  dissipationless transmission of 
information -- a prerequisite for quantum technologies. 
In this respect, applications of intense laser pulses hold potentials of signal processing at the Petahertz rate~\cite{boolakee2022light}. 
Thus  employing an ultrafast intense laser on novel quantum materials 
is an emerging avenue for converting light into electricity efficiently~\cite{bharti2023tailor,ikeda2023photocurrent,neufeld2021light}.
However, a universal method to transform light into electricity applicable to topological and normal materials in two and three dimensions is lacking. 

The present work introduces a universal way to generate photocurrent in topological and nontopological materials. 
We will start our discussion by  demonstrating photocurrent generation in 
inversion-symmetric and inversion-broken Weyl semimetals.  
Owing to a several picoseconds long electron scattering timescale, Weyl semimetals seem 
ideal for coherent light manipulation, including photocurrent generation ~\cite{ikeda2023light}. 
Analysis of the photocurrent in Weyl semimetals is also useful to unravel its topological aspects~\cite{hamara2023helicity,luo2021light,rees2020helicity,wang2019robust,chan2017photocurrents,gao2020chiral,osterhoudt2019colossal}. 
In addition, the plethora of interesting optical phenomena in Weyl semimetals makes them suitable candidates for 
interaction with intense laser~\cite{morimoto2016topological, bharti2023weyl, mciver2012control, ma2017direct,bharti2022high, bharti2023role, weber2021ultrafast}.
In recent years, two-color co- and counter-rotating circularly polarized laser pulses have been  employed to generate 
photocurrent in normal and  topological materials, respectively~\cite{neufeld2021light,ikeda2023photocurrent}. 
In addition, it has been demonstrated that a single-color circularly polarized pulse is useful to tailor photocurrent in Weyl  semimetals~\cite{bharti2023tailor}. 
However, a method based on linear polarization of light is desirable for its simplicity. 

A schematic setup of our idea is shown in Fig.~\ref{fig:fig1}(a) where a pair of  linearly polarized pulses with frequencies $\omega$ and $2\omega$ is shined on a Weyl semimetal. 
The polarization of the $\omega$ pulse is fixed along  the  $x$ axis,  
whereas  the polarization direction of 2$\omega$ is making an angle $\theta$ with respect to the $x$ axis in the $xy$ plane. 
When both $\omega-2\omega$ pulses are in collinear configuration ($\theta = 0$), 
the resultant laser pulse contains both the components, $\omega$ and 2$\omega$,
along the $x$ direction with no component along the $y$ direction. 
In addition, there are more oscillations on the negative side than the positive side resulting in an asymmetric laser waveform 
as evident from the top panel of Fig.~\ref{fig:fig1}(b). 
On the other hand, the orthogonal configuration of the $\omega-2\omega$ pulses, i.e., $\theta = \pi/2$, results in a symmetric laser waveform in both $x$ and $y$ directions as reflected from the bottom panel of Fig.~\ref{fig:fig1}(b). 
Note that the orthogonal configuration leads to an interesting Lissajous figure with reflection symmetry only along the $x$ direction as shown in the inset [see Fig.~\ref{fig:fig1}(b)].
Hence, orthogonal configuration results in an asymmetric laser waveform along the $y$ direction. 
Further, the asymmetry of the total laser waveform  can be tuned by changing the polarization direction and intensity of the $2\omega$ field. 

In the following, we will show the generation and the manipulation of the photocurrent in Weyl semimetals  when the setup is in collinear or  orthogonal configuration. 
Furthermore, it has been found that the presence of even a weak $2 \omega$ pulse is enough to generate photocurrent, 
which  can be further tailored by tuning the intensity of the $2 \omega$ pulse with respect to the $\omega$ pulse. 
In addition, the photocurrent can be further tuned by controlling the interplay of $\omega-2\omega$ pulses through variations in angle $\theta $, amplitude ratio, and time delay between them.
To demonstrate the universality and robustness of our idea, we will extend our study to inversion-symmetric and inversion-broken two-dimensional  materials with trivial topology.
Asymmetry of the carrier-envelope phase stabilized  few-cycle laser waveform has been utilized 
to generate photocurrent in  two-dimensional materials, where 
the photocurrent was controlled by tuning the carrier-envelope-phase of the pulse~\cite{higuchi2017light, heide2020sub, li2021ab}. 
In addition, coherent control over electronic motion and  current in graphene is achieved by 
a pair of few-cycle linearly polarized laser pulses~\cite{heide2018coherent, sharma2023thz}. 
Additionally, the resultant photocurrent gives a lower bound on coherence time in graphene~\cite{heide2021electronic}.
The present approach shows a universal application of intense light to engender photocurrent in both topological and non-topological  materials in two and three dimensions.
Notably, $\omega-2\omega$ pulses have been employed to explore strong-field driven light-matter interaction phenomena in solids, namely high-harmonic generation~\cite{vampa2015linking, liu2023valley, mrudul2021light, navarrete2020two, rana2022generation, luu2018observing, avetissian2022efficient}. It is well established that intense laser driven high-harmonic generation in solids is nonperturbative in nature~\cite{goulielmakis2022high, ghimire2019high}. Thus it is expected that the intense laser pulses will generate nonperturbative
photocurrent in solids.

\section{Theoretical Framework}
Interaction of an intense laser pulse with a solid is described within the density matrix framework. 
The temporal evolution  of the density matrix, $\rho$, can be written as~\cite{mrudul2021high, rana2022high}
\begin{equation}\label{eq2}
\dot{\rho}_{mn}^{\textbf{k}} = 
i \textbf{E}(t) \cdot \sum_l \left(\textbf{d}_{ml}^{~\textbf{k}_{t}}\rho_{ln}^{\textbf{k}} - \textbf{d}_{ln}^{~\textbf{k}_{t}}\rho_{ml}^{\textbf{k}}  \right) - \left[ \frac{(1-\delta_{mn})}{\textrm{T}_{2}} + i \mathcal{E}_{mn}^{\textbf{k}_{t}} \right]\rho_{mn}^{\textbf{k}}.
\end{equation}
In the above equation, $\textbf{A}(t)$ and $\textbf{E}(t)$ are, respectively, vector potential and electric field of the incident laser pulse, which are related as $\textbf{E}(t) = - \dot{\textbf{A}}(t)$. 
In the presence of the laser pulse, crystal momentum $\textbf{k}$ changes to $\textbf{k}_{t} = \textbf{k}+\textbf{A}(t)$. 
In this work, the electronic band structure of solids are described by 
a tight-binding Hamiltonian $\mathcal{H}(\textbf{k})$, 
which is diagonalized to obtain eigenstates $|m,\textbf{k}\rangle$ and $|n,\textbf{k}\rangle$ corresponding to eigenvalues $\mathcal{E}^{\textbf{k}}_{m}$ and $\mathcal{E}^{\textbf{k}}_{n}$, respectively, at each time step in the presence of laser. 
After obtaining the eigenstates, 
the momentum and dipole  matrix elements are, respectively,  
calculated as  
$\textbf{p}_{mn}^{\textbf{k}} = \langle m,\textbf{k}|\nabla_\textbf{\textbf{k}}\mathcal{H}(\textbf{k})| n,\textbf{k}\rangle$ and 
$\textbf{d}_{mn}^{\textbf{k}} = i\langle m,\textbf{k} |\nabla_\textbf{k}|n,\textbf{k}\rangle$ 
at each time step numerically as well as the energy gap, $\mathcal{E}^{\textbf{k}}_{mn} = \mathcal{E}^{\textbf{k}}_m - \mathcal{E}^{\textbf{k}}_n$.
A phenomenological dephasing time $\textrm{T}_2 = $ 1.5 fs is introduced to consider decoherence between electron and hole during the excitation process.

Equation \eqref{eq2} is solved numerically by sampling  the Brillouin zone with $80 \times 80 \times 80$ grid size 
and a time step of 0.015 fs.  
We start with a filled valence band and an empty conduction band.
Using this as the initial condition, Eq. \eqref{eq2} is solved using the Runge-Kutta method to obtain the density matrix, $\rho(t)$, at each time $t$.
A nonzero photocurrent,  $\mathbf{J}(t) = \int_\mathbf{k}  d\mathbf{k}~\left[ \rho(\mathbf{k}) - \rho(-\mathbf{k}) \right] \pdv{\mathcal{E}(\mathbf{k})}{\mathbf{k}}$,  arises if there is an asymmetric electronic population in the conduction band 
after the end of the laser pulse, i.e., $\rho(\mathbf{k}) \neq \rho(-\mathbf{k})$~\cite{bharti2023tailor,soifer2019band}. In the present work, we analyze the photocurrent for various configurations of the laser pulse to ascertain suitable configurations for producing the asymmetric population and hence, in turn, photocurrent in solids.

\begin{figure}
\centering
\includegraphics[width=\linewidth]{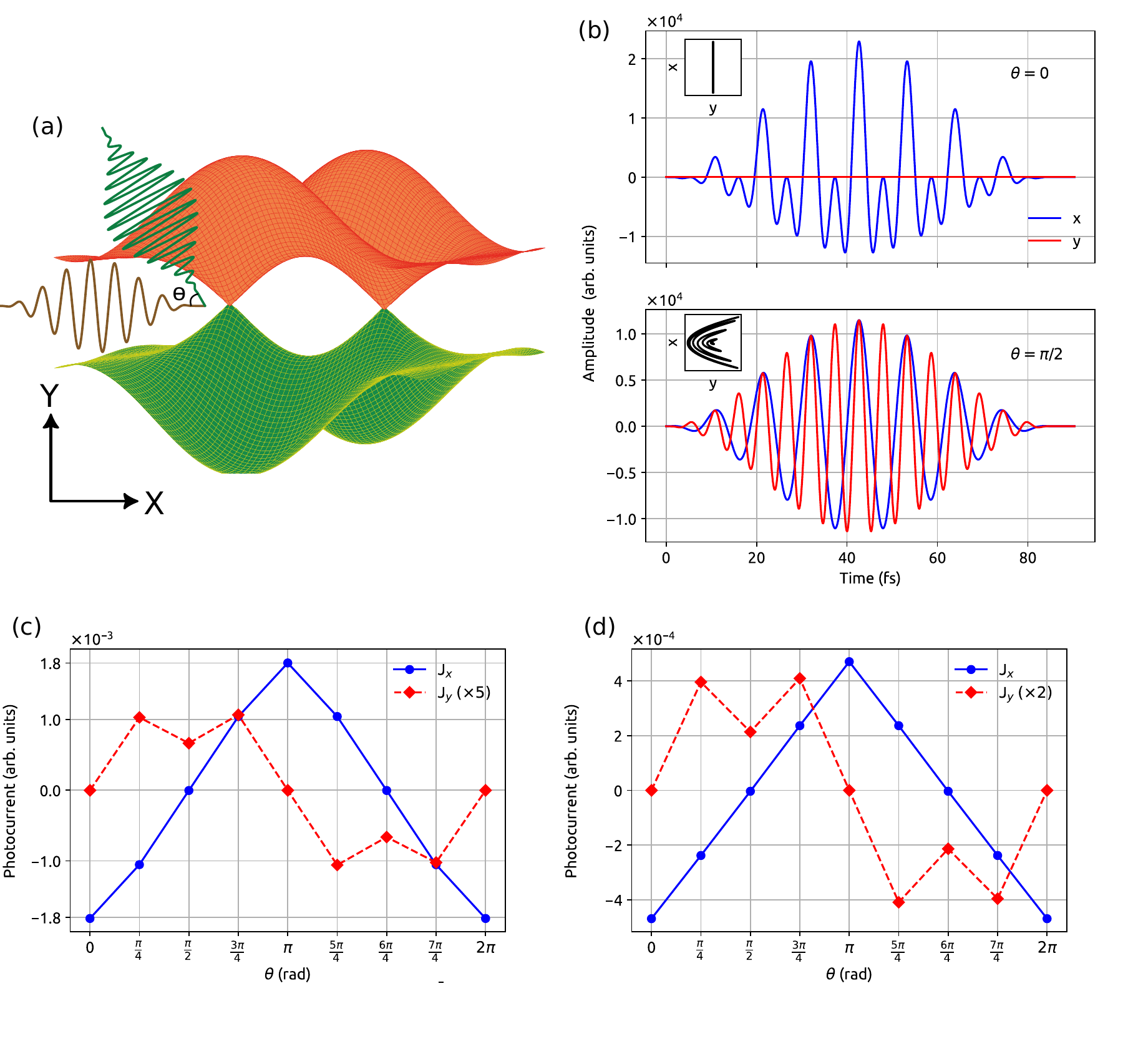}
\caption{(a) Schematic setup for ultrafast photocurrent generation where two linearly polarized pulses with frequencies 
$\omega$ and $2 \omega$ with polarization directions at an angle $\theta$ are interacting with a Weyl semimetal. 
(b)  Amplitude of the vector potential of the two pulses when both pulses are in collinear (top panel) and orthogonal (bottom panel) configurations. 
The Lissajous curve of the total vector potential in the $xy$ plane is shown in respective insets. 
(c) Variations in the photocurrent with respect to $\theta$ in an inversion-symmetric Weyl semimetal. (d) Same as (c) for
an inversion-broken Weyl semimetal. 
Wavelength of the $\omega$ pulse is 3.2 $\mu$m with pulse length $\sim$ 100 fs. 
Laser intensity equal to $5\times10^{10}$ W/cm$^2$  with $\mathcal{R}  = 1$ is used for both 
inversion-symmetric and inversion-broken Weyl semimetals.} \label{fig:fig1}
\end{figure}

The total vector potential of a pair of linearly polarized laser pulses is written as 
\begin{equation}\label{eq:laser}
\textbf{A}(t) = A_0 f(t)   \left[ \cos(\omega t) \hat{e}_{x} + \mathcal{R} \cos(2\omega t) \{ \cos(\theta) \hat{e}_{x}  + 
\sin(\theta) \hat{e}_y  \} \right],      
\end{equation}
where $A_0$ is the amplitude, $f(t)$ is the $\sin^2$ envelope,  
$\mathcal{R}$ is a dimensionless parameter to tune the intensity of the $2\omega$ pulse with respect to the $\omega$ pulse 
and  $\theta$ is the angle between the polarization  of two linearly polarized pulses as shown in Fig.~\ref{fig:fig1}(a). 

Let us write the Hamiltonian of Weyl semimetals as 
$\mathcal{H}(\textbf{k}) = \textbf{d}(\textbf{k}) \cdot \sigma$, where $\sigma$'s are the Pauli matrices. 
The components of $\textbf{d}(\textbf{k})$ for an inversion-broken Weyl semimetal are expressed as~\cite{sadhukhan2021role}  
\begin{eqnarray}\label{eq:invb}
	&\textbf{d}(\textbf{k})  =  & \left[t\{\cos(k_0 a)-\cos(k_y a) + \mu[1-\cos(k_z a)]\}, t\sin(k_z a), \right. \nonumber \\
	&& \left. t\{\cos(k_0 a)-\cos(k_x a)+\mu[1-\cos(k_z a)]\}\right],
\end{eqnarray}
and for an inversion-symmetric Weyl semimetal as 
\begin{equation}\label{eq:trb}
\textbf{d}(\textbf{k})  =   \left[t\sin(k_x a), t\sin(k_y a),t\{\cos(k_z a) - \cos(k_0 a)
 +2- \cos(k_x a) - \cos(k_y a)\}\right].
\end{equation}
Here, $k_{0} = \pi/(2a) $ represents the position of the Weyl nodes for both Weyl semimetals  with 
$a = 6.28$ \AA~as the lattice parameter of a simple cubic crystal structure. 
The Weyl nodes for inversion-broken and inversion-symmetric Weyl semimetals are at 
$\textbf{k}=[\pm\pi/(2a),\pm\pi/(2a),0]$ and $\textbf{k} = [0,0,\pm \pi/(2a)]$, respectively. 
A dimensionless parameter $\mu = 2$ and isotropic hopping parameter $t=1.8$ eV are used in the present work.
Similar tight-binding Hamiltonians for pristine graphene and  MoS$_2$ are adopted from 
Refs.~\cite{mrudul2021high} and ~\cite{rana2023all}, respectively. 

\section{Results and Discussion}
\subsection{Photocurrent in Weyl Semimetals}
We start our discussion by analyzing the sensitivity of the photocurrent with respect to $\theta$ for an inversion-symmetric Weyl semimetal as shown in Fig.~\ref{fig:fig1}(c). 
It is evident that there is nonzero photocurrent along the laser polarization for $\theta = n \pi$ with $n$ as an  integer. 
The origin of the finite photocurrent can be attributed to the asymmetric laser waveform [see Fig.~\ref{fig:fig1}(b)], 
which results in an asymmetric residual electronic population in the conduction band, i.e.,  $\left[ \rho(\mathbf{k}) - \rho(-\mathbf{k}) \right] \neq 0$. 
The asymmetry of the laser waveform can be flipped by changing  $\theta = 0$ or $2 \pi$ to $\pi$. 
As a result,  the magnitude of the photocurrent tunes from negative to positive by changing the collinear configuration from parallel to antiparallel.
In addition, the waveform asymmetry reduces as  $\theta$ deviates from the collinear configuration, such as at 
$\theta=\pi/4$, which leads to the reduction of the photocurrent. 
The orthogonal configuration of the $\omega - 2 \omega$ pulses renders 
a symmetric waveform along the $x$ direction, which results in zero photocurrent. 
Thus the underlying mechanism for the photocurrent generation is the asymmetric laser waveform, which  is imprinted in the excitation processes leading to asymmetric electronic  population.
Interestingly, there is also a small photocurrent along the $y$ direction for $\theta \neq n \pi$ as reflected in the figure.
The photocurrent along the $y$ direction arises  due to the asymmetric waveform  in the $y$ direction [see inset of bottom panel of Fig.~\ref{fig:fig1}(b)].
We will  discuss later how this photocurrent can be enhanced.
Thus the analysis of Fig.~\ref{fig:fig1}(c) establishes that the magnitude and direction of the 
photocurrent is tunable with  $\theta$.

Figure~\ref{fig:fig1}(d) presents  the variation in the photocurrent with $\theta$ for an inversion-broken 
Weyl semimetal. 
The collinear configuration results in a finite photocurrent in this case also. 
This observation, together with the sensitivity of the photocurrent's magnitude for other $\theta$, 
exhibits similar behavior as in the case of the inversion-symmetric Weyl semimetal discussed above.
In addition, the trend in the photocurrent is universal in the sense that it does not depend on the inversion symmetry of the material.
Thus Figs.~\ref{fig:fig1}(c) and (d) establish that the $\omega - 2 \omega$ pulse setup 
generates a finite photocurrent, which emanates by imprinting asymmetry of laser waveform on the electronic  population. 
The overall magnitude of the photocurrent is tunable with  $\theta$ and critically depends on the material and intensity employed.
So far, we have investigated photocurrent generation within the $\omega - 2 \omega$ setup with identical intensity of the laser pulses. 
At this junction, it is worth wondering how the photocurrent changes when the intensity ratio of the two pulses varies. 

\begin{figure}
\centering
\includegraphics[width=\linewidth]{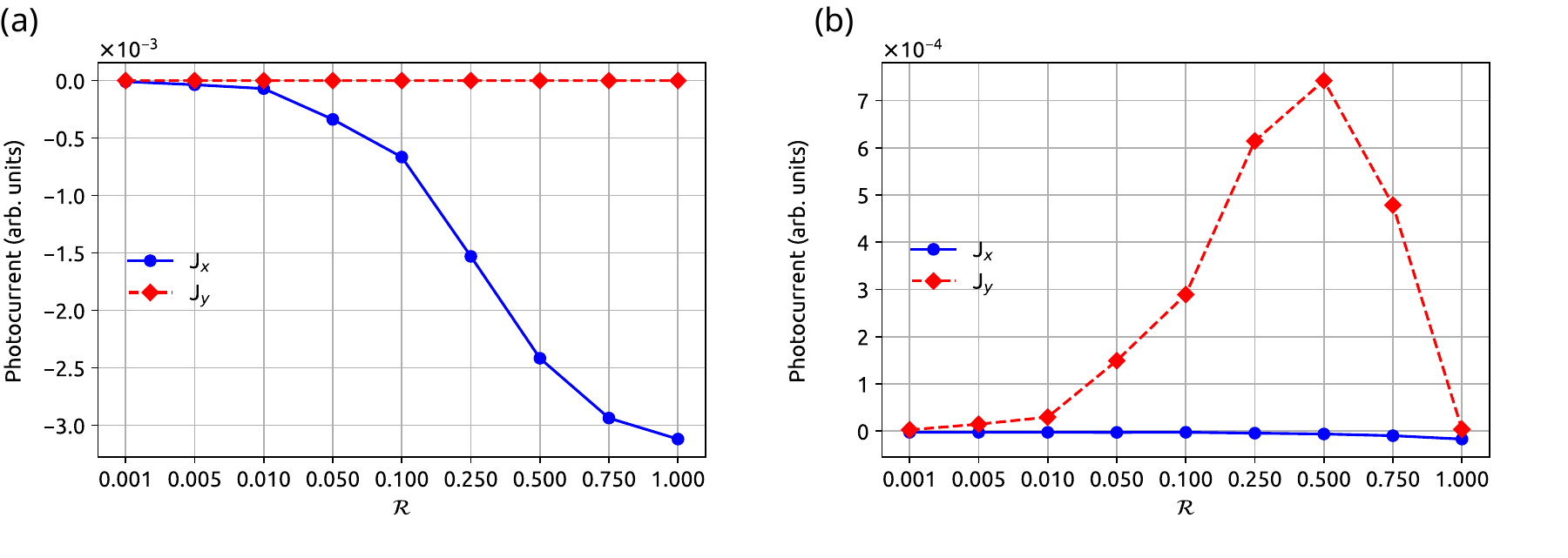}
\caption{Variation in photocurrent with the amplitude ratio  ($ \mathcal{R}$) 
of the $2\omega$ pulse with respect to the $\omega$ pulse for an 
inversion-symmetric Weyl semimetal in the  (a) collinear and (b) orthogonal configurations. 
Laser parameters are the same as in Fig.~\ref{fig:fig1} with intensity of the $\omega$ pulse as $10^{11}$ W/cm$^2$.} \label{fig:fig2}
\end{figure}

\subsection{Role of the Intensity Ratio ($\mathcal{R}$) in  Photocurrent}
To address this pertinent issue, we will consider two  cases where the waveform's asymmetry 
is extremum, i.e., $\theta = 0$  (collinear) and $\pi/2$ (orthogonal) configurations. 
Moreover, an inversion-symmetric Weyl semimetal is chosen for further discussion from here onward as both inversion-symmetric and inversion-broken Weyl semimetals behave similarly. 
Figure~\ref{fig:fig2} presents the sensitivity of the photocurrent as a function of the amplitude ratio  
$\mathcal{R}$ [see Eq.~\eqref{eq:laser}]. 
The residual electronic population in the conduction band, after the end of the laser pulse,  
exhibits asymmetry along $k_x=0$ and the asymmetry increase with  $\mathcal{R}$ [see Fig. S1~\cite{NoteX}]. 
Consequently, the photocurrent becomes significant as $\mathcal{R}$ is increased, as evident from Fig.~\ref{fig:fig2}(a). 
Further, it is notable that the presence of a weaker $2 \omega$ pulse is enough to
generate photocurrent of the same order of magnitude.
Also, in comparison  to $\mathcal{R} = 1$, there is an appreciable photocurrent even when the intensity of the 
$2\omega$ field is  one-tenth of the $\omega$ field. 
In general, the generation of the $2\omega$  pulse from $\omega$, say using a beta barium borate crystal, reduces the intensity of the $2\omega$  pulse drastically in typical experimental setups. 
Thus the presence of a 
weaker $2\omega$ pulse in the $\omega - 2\omega$  setup in our approach 
is sufficient for the photocurrent generation to provide flexibility to the experimentalist.
Note that the spatial distribution of the residual population resembles the laser waveform as  
crystal momentum $\mathbf{k}$ alters to $\mathbf{k}_{t} = \mathbf{k} + \mathbf{A}(t)$.  
Thus, the laser waveform controls the  asymmetry of  the residual  population and therefore photocurrent.

\begin{figure}
\centering
\includegraphics[width=\linewidth]{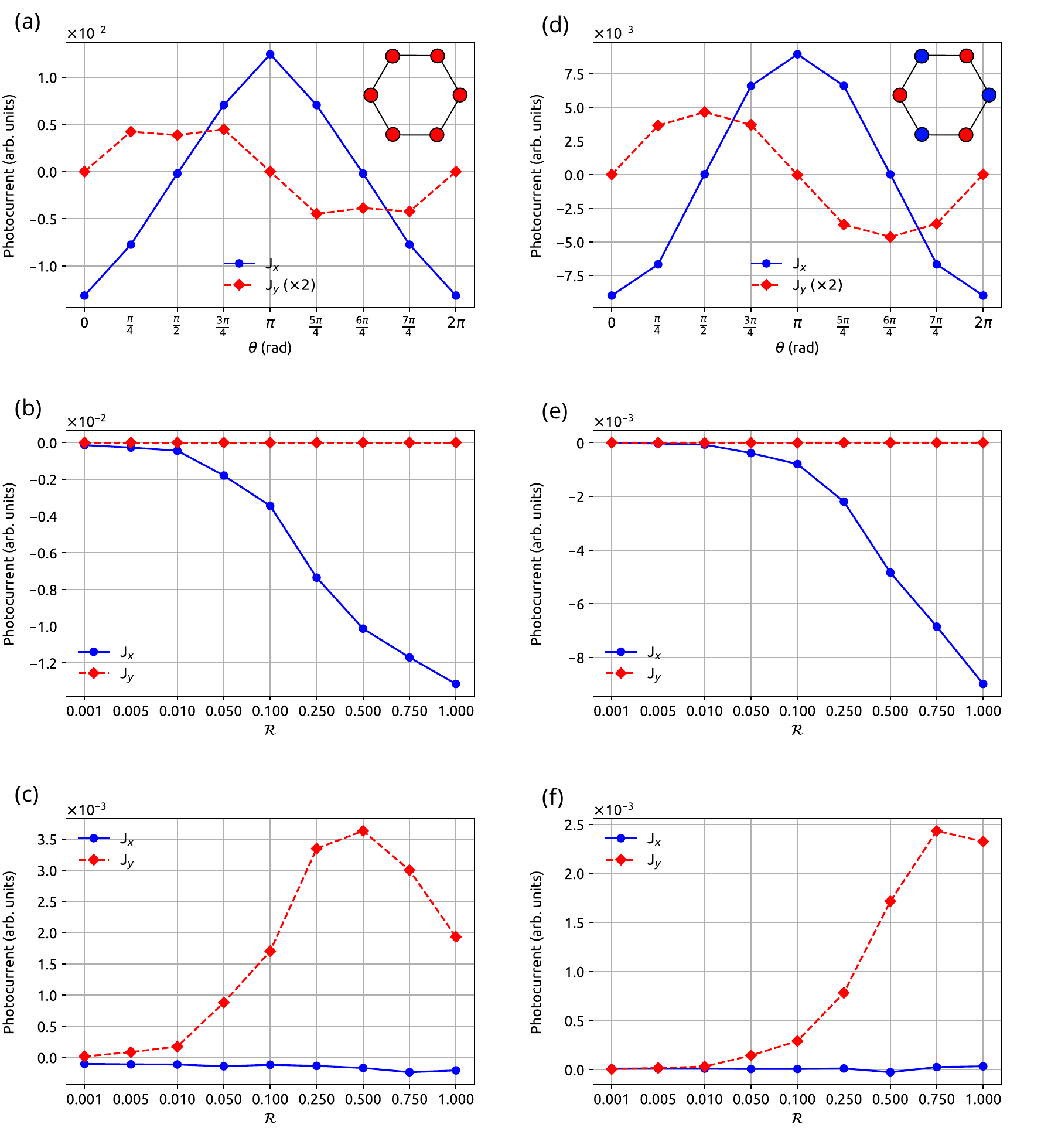}
\caption{(a) Sensitivity of  the photocurrent in an inversion-symmetric graphene with respect to 
(a) $\theta$, the amplitude ratio ($ \mathcal{R}$)  in (b) collinear and (c) orthogonal configurations of the $\omega-2\omega$ setup. Panels (d), (e), and (f) are the same as (a), (b), and (c) for MoS$_2$, respectively. 
Laser parameters are the same as in Fig.~\ref{fig:fig1} with intensity of the $\omega$ pulse as $10^{11}$ W/cm$^2$.}
\label{fig:fig3}
\end{figure}

Analysis of Fig.~\ref{fig:fig1}(c) indicates that there is insignificant photocurrent along the $y$ direction  
when $\theta = \pi/2$ for $\mathcal{R} = 1$.
However, photocurrent along the $y$ direction can be boosted by an order of magnitude by tuning $\mathcal{R} = 1$ to 0.5 as reflected from Fig.~\ref{fig:fig2}(b). 
Similar to the collinear configuration, the residual population in the 
conduction band is fairly symmetric along $k_x = 0$ for $\mathcal{R}=0.1$. 
On the other hand, the population is asymmetric along $k_y=0$, which results in photocurrent along the $y$ direction 
[see Fig. S2~\cite{NoteX}].
Moreover, the orthogonal configuration exhibits the non-monotonic behavior of the photocurrent, which is in contrast to the observation in the collinear configuration.
The nonmonotonic behavior can be attributed to the laser-driven  nonperturbative  electron dynamics in the conduction bands. 
A further increase in the intensity, by increasing $\mathcal{R}$, leads to the sign change of $\left[\rho(\mathbf{k})-\rho(-\mathbf{k})\right]$, which results in the reversal of the photocurrent's direction. Our observation about the direction reversal with intensity is 
consistent with previous reports~\cite{bharti2023tailor,zhang2022bidirectional,higuchi2017light,wachter2015controlling,wismer2016strong}.
The same observations hold true for an inversion-broken Weyl semimetal qualitatively.  
Thus, $\mathcal{R}$ adds another knob to tune the photocurrent in Weyl semimetals along with $\theta$. 

\subsection{Photocurrent in Two-Dimensional Materials} 
Until now, we have observed that the photocurrent is insensitive to the symmetries and topology  
of the Weyl semimetals 
and exhibits similarities with $\theta$  and $\mathcal{R}$ variations. 
This conclusion raises a crucial question about the universality of our observation. 
To answer this important question, we transit from three-dimensional topological to two-dimensional trivial materials, namely inversion-symmetric graphene and inversion-broken molybdenum disulfide (MoS$_2$). 
These two-dimensional materials have been the center of exploration for photodetection and other optoelectronic applications in recent years~\cite{mrudul2021controlling,rana2023all,koppens2014photodetectors,agarwal_2023, mak2016photonics, bussolotti2018roadmap}. 

There is a finite photocurrent along the $x$ direction for graphene in the collinear configuration 
($\theta = 0, \pi$ and $2 \pi$)  as evident from Fig.~\ref{fig:fig3}(a). 
On the other hand, the photocurrent's magnitude reduces as $\theta$ changes and reaches a minimum  
for  $\theta = \pi/2$ and $3\pi/2$. 
A similar trend in the photocurrent with $\theta$  is visible for MoS$_2$ [see Fig.~\ref{fig:fig3}(d)].
Interestingly, the photocurrent's magnitude in MoS$_2$ is reduced in comparison to graphene, 
which can be attributed to the finite band gap of MoS$_2$.
Overall, other features of the photocurrent along $x$ and $y$ directions remain robust with a variation in $\theta$. 

\begin{figure}
\centering
\includegraphics[width=\linewidth]{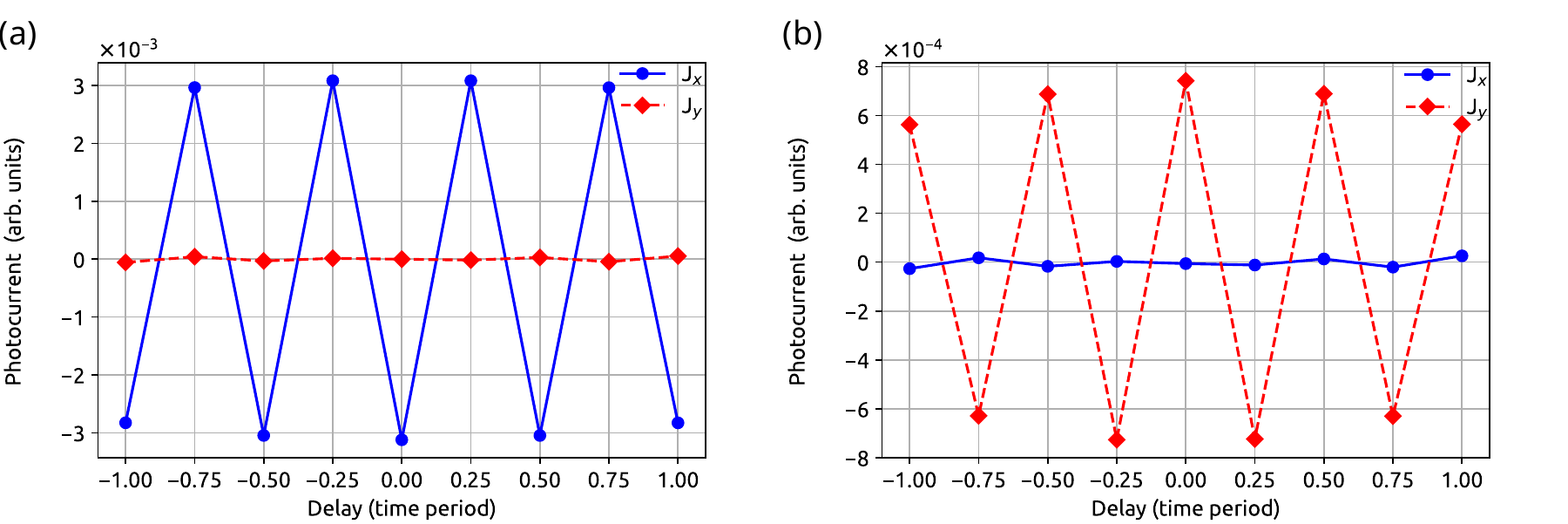}
\caption{Effect of the time-delay between  of $\omega$ and $2\omega$ pulses on the photocurrent in  
an inversion-symmetric Weyl semimetal. 
The $\omega$ and $2\omega$ pulses are in (a) collinear and (b) orthogonal configurations 
with $ \mathcal{R} = 1$ and 0.5, respectively. 
The laser parameters are the same as in Fig.~\ref{fig:fig1}.} \label{fig:fig4}
\end{figure}

Sensitivity of the photocurrent with $\mathcal{R}$ 
is presented in Figs.~\ref{fig:fig3}(b) and ~\ref{fig:fig3}(e) for graphene and MoS$_2$, respectively. 
It is noted that the small presence of the $2 \omega$ component in collinear configuration is sufficient to generate photocurrent along the $x$ direction. 
On the other hand, the value of $\mathcal{R}$ crucially depends on the material's nature to maximize the photocurrent along the $y$ direction in the orthogonal configuration, as evident 
from Figs.~\ref{fig:fig3}(c) and ~\ref{fig:fig3}(f) for graphene and MoS$_2$, respectively.  
Observations of Figs.~\ref{fig:fig1} - \ref{fig:fig3} confirm  that the  photocurrent can be tuned by varying  th elaser's parameters in the $\omega-2\omega$ setup irrespective of the materials and their underlying symmetry, which  
establishes the universality of our approach.

\subsection{Role of Time-Delay between $ \omega - 2 \omega$  Pulses in Photocurrent} 
At this juncture, we investigate how the photocurrent can be additionally controlled by introducing a relative time delay between $\omega$ and $2 \omega$ pulses. So far, we have considered zero delays between the pulses.  
Figure~\ref{fig:fig4}(a) presents the variation in the photocurrent as a function of the delay for an inversion-symmetric Weyl semimetal in a collinear configuration ($\theta = 0$) 
with $\mathcal{R} = 1$ at which the photocurrent is maximum [see Fig.~\ref{fig:fig2}(a)].  
The photocurrent's amplitude along the $x$ direction can be modulated from negative to positive value by changing the relative delay in units of a quarter of the fundamental ($\omega$ pulse) time period. 
The modulation in the photocurrent can be attributed to the change in the asymmetry of the laser waveform caused by the time delay.  
Figure~\ref{fig:fig4}(b) shows a similar control over the photocurrent's amplitude along the $y$ direction  
in the  orthogonal  configuration ($\theta = \pi/2$) with $\mathcal{R} = 0.5$ at which the photocurrent is maximum [see Fig.~\ref{fig:fig2}(b)].
Thus the photocurrent can be modulated by merely introducing the delay, which adds another convenient control knob to tailor photocurrent in materials.   

\begin{figure}
\centering
\includegraphics[width=\linewidth]{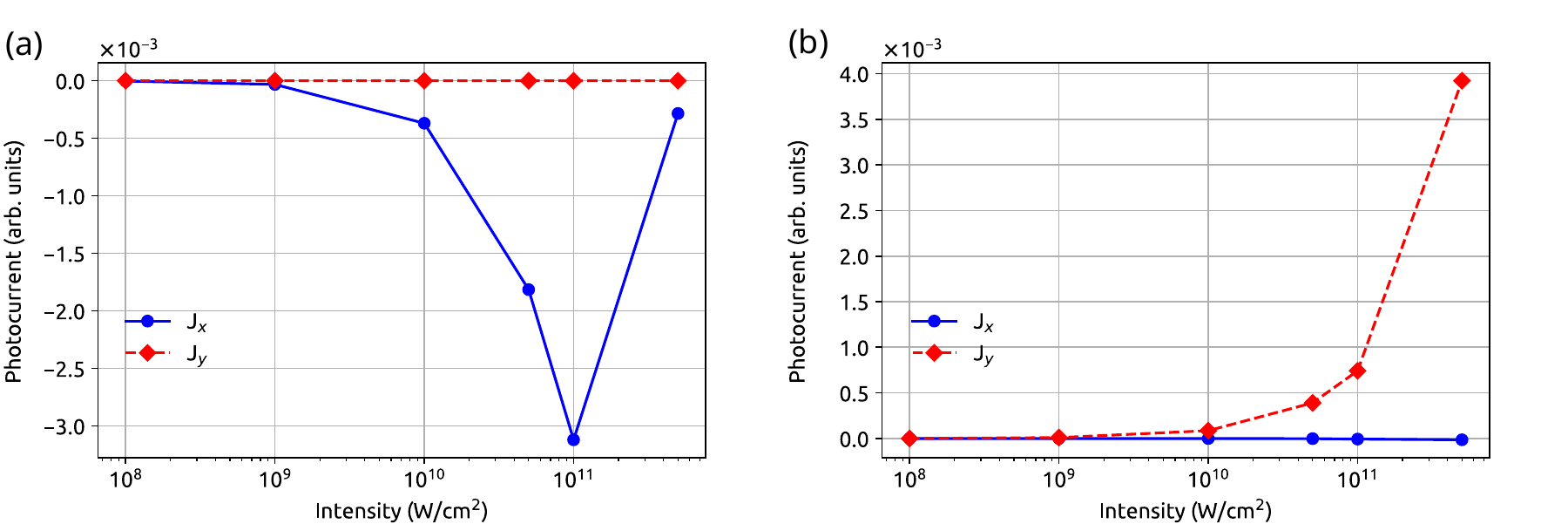}
\caption{Scaling of the photocurrent  with the intensity of the 
$\omega$ and $2\omega$ pulses in (a) collinear and (b) orthogonal configurations 
with $ \mathcal{R} = 1$ and 0.5, respectively. 
The laser parameters are the same as in Fig.~\ref{fig:fig1} for an inversion-symmetric Weyl semimetal.}
\label{fig:fig5}
\end{figure}

\subsection{Role of the Laser's Intensity in Photocurrent} 
So far, we have limited our discussion on the photocurrent for a fixed intensity of the $\omega$ pulse.  
At this point, it is worth knowing how photocurrent scales with the intensity.
Figure~\ref{fig:fig5} discusses how the photocurrent scales with the intensity   
in collinear and orthogonal configurations. 
The intensity of both $\omega$ and $2\omega$ pulses 
are varied in a fixed ratio, which corresponds to the maximum photocurrent as in Fig.~\ref{fig:fig2}.
It is evident that the photocurrent becomes appreciable at $10^{10}$ W/cm$^2$ and exhibits nonmonotonic behavior 
in the collinear configuration as shown in Fig.~\ref{fig:fig5}(a). 
The photocurrent peaks at $10^{11}$ W/cm$^2$ and starts decreasing with an increase in intensity, which  
results in the reversal of the photocurrent's direction as discussed above. 
The asymmetry in the residual electronic population increases along $k_{x}$ as the intensity increases, which leads to increase in photocurrent. However, after reaching  a maximum, the asymmetry starts reducing as the    
 residual  population migrates from positive  to negative $k_{x}$ region and vice versa, 
 which results in the reduction of the photocurrent's magnitude, and can be understood by analyzing
the residual population as shown in Fig. S3~\cite{NoteX}. 
The  minimum intensity required to generate photocurrent  and its nonmonotonic nature  indicate that the generated  photocurrent is nonperturbative in nature. 
This observation is consistent with an earlier report for graphene exposed to a few-cycle phase-stabilized laser pulse~\cite{higuchi2017light,zhang2022bidirectional} and Weyl semimetal~\cite{bharti2023tailor}.
In contrast,  the orthogonal configuration at the same intensity yields minuscule photocurrent as reflected from Fig.~\ref{fig:fig5}(b). 
Photocurrent in the orthogonal  configuration increases monotonically with the laser's intensity studied.
 The residual population in the conduction band  in the orthogonal configuration for different laser's intensity   is presented in  Fig. S4~\cite{NoteX}, which can be analyzed in a similar fashion as discussed above.
Note that the maxima and directional reversal of photocurrent will appear at intensity different from Fig.~\ref{fig:fig5}, 
if we choose a different value of $\mathcal{R}$ for any configuration.
Nonetheless, above a threshold intensity, the $\omega-2\omega$ field can produce photocurrent which can be optimized by tuning the ratio of amplitude.

\section{Conclusion}
In summary, we present a universal method to generate and tailor photocurrent in normal and topological materials, namely graphene, MoS$_2$ and Weyl semimetals. 
A pair of linearly polarized pulses comprised of $\omega-2\omega$ frequencies  can produce a highly asymmetric laser 
waveform, which steers electrons on an attosecond timescale to generate photocurrent in materials with trivial and 
nontrivial topology.
In this regard, we find that the presence of a comparatively weak $2\omega$ pulse is sufficient for the photocurrent generation.
Interestingly, the generated photocurrent can be tailored by simply varying the laser parameters of the $\omega-2\omega$ setup. 
The photocurrent is found to be sensitive to the variation in the angle between the polarization directions, amplitude ratio, and relative time delay of the two pulses.
Even orthogonal linearly polarized pulses drive asymmetric population for a certain amplitude ratio and thus give 
rise to comparable photocurrent as the collinear pulses. 
Our proposed method showcases various ways to tailor laser waveform to generate photocurrent for optoelectronic and photodetection applications in a nonmaterial specific manner -- thus a universal way.

\section*{Acknowledgment}

G. D. acknowledges support from SERB India  (Project No. MTR/2021/000138).   

%
\end{document}